\documentclass{article}
\usepackage{arxiv}
\usepackage[utf8]{inputenc} % allow utf-8 input
\usepackage[T1]{fontenc}    % use 8-bit T1 fonts
\usepackage{hyperref}       % hyperlinks
\usepackage{url}            % simple URL typesetting
\usepackage{booktabs}       % professional-quality tables
\usepackage{amsfonts}       % blackboard math symbols
\usepackage{nicefrac}       % compact symbols for 1/2, etc.
\usepackage{microtype}      % microtypography
\usepackage{lipsum}
\usepackage{hyperref}
\usepackage{amsmath,amsfonts,amsthm,amssymb,bm,bbm} % Math packages
\usepackage{MnSymbol}

\newcommand{\R}{\mathbb{R}}
\makeatother
\usepackage{dirtytalk}
\usepackage{float}
\usepackage{siunitx}
\usepackage{booktabs}
\usepackage{multirow}
\usepackage{tabulary}
\usepackage{graphicx}

\title{AttentionDDI: Siamese Attention-based Deep Learning method for drug-drug interaction predictions}

\author{
  Kyriakos Schwarz\thanks{Biomedical Informatics, University Hospital of Zurich, Zurich, Switzerland} \\
  Department of Quantitative Biomedicine\\
  University of Zurich\\
  Schmelzbergstrasse 26, Zurich, CH \\
  \texttt{kyriakos.schwarz@uzh.ch} \\
   \And
  Ahmed Allam\footnotemark[1] \\
  Department of Quantitative Biomedicine\\
  University of Zurich\\
  Schmelzbergstrasse 26, Zurich, CH \\
  \texttt{ahmed.allam@uzh.ch} \\
   \And
    Nicolas Andres Perez Gonzalez\footnotemark[1] \\
  Department of Quantitative Biomedicine\\
  University of Zurich\\
  Schmelzbergstrasse 26, Zurich, CH \\
  \texttt{nicolas.perezgonzalez@uzh.ch} \\
   \And
  Michael Krauthammer\footnotemark[1] \\
  Department of Quantitative Biomedicine\\
  University of Zurich\\
  Schmelzbergstrasse 26, Zurich, CH \\
  \texttt{michael.krauthammer@uzh.ch} \\
}

\begin{document}
\maketitle

\begin{abstract} % abstract

\textbf{Background} %if any
Drug-drug interactions (DDIs) refer to processes triggered by the administration of two or more drugs leading to side effects beyond those observed when drugs are administered by themselves. Due to the massive number of possible drug pairs, it is nearly impossible to experimentally test all combinations and discover previously unobserved side effects. Therefore, machine learning based methods are being used to address this issue.

\textbf{Methods} %if any
We propose a Siamese \textit{self-attention} multi-modal neural network for DDI prediction that integrates multiple drug similarity measures that have been  derived from a comparison of drug characteristics including drug targets, pathways and gene expression profiles.

\textbf{Results}
Our proposed DDI prediction model provides multiple advantages: 1) It is trained end-to-end, overcoming limitations of models composed of multiple separate steps, 2) it offers model explainability via an \textit{Attention} mechanism for identifying salient input features and 3) it achieves similar or better prediction performance (AUPR scores ranging from $0.77$ to $0.92$) compared to state-of-the-art DDI models when tested on various benchmark datasets. Novel DDI predictions are further validated using independent data resources. 

\textbf{Conclusions}
We find that a Siamese multi-modal neural network  is able to accurately predict DDIs and that an \textit{Attention} mechanism, typically used in the Natural Language Processing domain,  can be beneficially applied to aid in DDI model explainability.

\end{abstract}

% keywords can be removed
\vspace*{0.5cm}

\clearpage
\section*{Introduction}
%\subsection*{Background}
%\paragraph*{} 
Polypharmacy, the concurrent administration of multiple drugs, has been increasing among patients in recent years \cite{kantor_trends_2015}. When administering multiple drugs, interactions might arise among them, often termed drug-drug interactions (\textbf{DDI}). The intended effect of a drug may therefore be altered by the action of another drug. These effects could lead to drug synergy, reduced efficacy or even to toxicity. Thus, DDI discovery is an important step towards improved patient treatment and safety.

%\paragraph*{} 
It is almost impossible to conduct an empirical assessment of all possible drug pair combinations and test their propensity for triggering DDIs. Computational approaches have addressed this issue by enabling the testing of large number of drug pairs more efficiently. For instance, \textit{DeepDDI} \cite{ryu_deep_2018}, a multilabel classification model, takes drug structure data as input along with drug names, in order to make DDI predictions in the form of human-readable sentences. Another model, \textit{GENN} \cite{ma_genn_2019}, a graph energy neural network, puts a focus on DDI types and estimates correlations between them. \textit{NDD} \cite{rohani_drug-drug_2019} utilizes multiple drug similarity matrices, which are combined by Similarity Network Fusion (\textit{SNF}) and finally fed through a feed-forward network for classification. Similarly, ISCMF \cite{rohani_iscmf_2020} performs matrix factorization on the known DDIs in order to calculate latent matrices which are used for predictions. It utilizes the same \textit{SNF}-fused matrix as to constrain this factorization.

%\paragraph*{} 
The above mentioned solutions come with some drawbacks. First, there is a plethora of drug feature information available for many approved drugs, including chemical structure, side effects, targets, pathways, and more. However, current DDI prediction solutions often only take advantage of a small subset of these features, particularly drug chemical structure features, due to their broad availability. Other current model limitations include low interpretability and/or the fact that they consist of multiple separate steps (i.e., cannot be trained end-to-end). A novel solution should preferably offer a mechanism to tackle those drawbacks simultaneously.

%\paragraph*{} 
To this end, we introduce \textbf{AttentionDDI}, a Siamese \textit{self-attention}  multi-modal neural network model for DDI prediction. Our model is inspired by and adapts ideas from Attention-based models (i.e., Transformer network) \cite{vaswani_attention_2017} that showed great success particularly in the Natural Language Processing (\textit{NLP}) domain. Our model 1) is trained end-to-end, 2) offers model explainability and 3) achieves similar or better prediction performance compared to state-of-the-art DDI models when tested on various benchmark datasets.

\section*{Methods}
\subsection*{Benchmark datasets}
In order to predict interactions between drugs, we focused on specific benchmark datasets listed in Table \ref{table:benchmark_datasets}. Our model, \textit{AttentionDDI}, and two competitive  baseline models (developed by the same group), \textit{NDD} \cite{rohani_drug-drug_2019} and \textit{ISCMF} \cite{rohani_iscmf_2020}, are all built to take advantage of the multi-modality contained in those datasets. Each dataset consists of one or more drug similarity matrices as described in Table \ref{table:benchmark_datasets}. Those matrices are calculated based on various drug feature vectors, encoding diverse drug characteristics, including, for example, the side effects that are induced by a drug or the biological pathways a drug is targeting. All together,  the drug similarity matrices are based on the following drug characteristics: chemical structure, targets, pathways, transporter, enzyme, ligand, indication, side effects, offside effects, GO terms, PPI distance, and ATC codes. The datasets have been previously used by multiple other studies \cite{zhang_predicting_2017, wan_neodti_2019,gottlieb_indi_2012, rohani_drug-drug_2019, rohani_iscmf_2020}.

Additionally to the above mentioned matrices, we calculate the Gaussian Interaction Profile (GIP) similarity matrix (according to \cite{van_laarhoven_gaussian_2011}) based on the interaction labels of each dataset (Table \ref{table:labels}). Therefore, in addition to the similarity features listed in Table \ref{table:benchmark_datasets}, the GIP of each dataset label matrix is also utilized as a further similarity feature. This method works under the hypothesis that drugs with resembling existing labels (DDIs) are expected to have comparable novel interaction predictions.

\begin{table}[ht]
\begin{center}
    \normalsize
\begin{tabular}{l|r|l}
Dataset & $\#$ drugs & Similarity Matrices                                                                      \\\hline
DS1 \cite{zhang_predicting_2017}     & 548 &Chemical, Enzyme, Indication, Offside effects, Pathway, Side effects, Target, Transporter \\
DS2 \cite{wan_neodti_2019}    & 707    & Chemical                                                                                 \\
DS3 \cite{gottlieb_indi_2012}     & 807  & ATC, Chemical, GO, Ligand, PPI Distance, Side effects, Target                            
\end{tabular}
    \end{center}
    \caption{Benchmark datasets.\label{table:benchmark_datasets}}
\end{table}

We obtained the precomputed drug similarity matrices from \cite{rohani_drug-drug_2019}. As an example, the \textit{side effects} matrix of the DS1 dataset \cite{zhang_predicting_2017} was constructed as follows: A matrix representing a list of $N$ known drugs on the $y$-axis and a list of $M$ known side effects on the $x$-axis was created. In this matrix, each row is representing a drug along with its side effects in the $N \times M$ matrix. It is filled with the value $1$ in each position where it is known that a drug may cause a specific side effect, $0$ otherwise. In this fashion, each drug is represented by a binary feature vector (size $M$). Furthermore, this binary feature matrix was transformed into a similarity matrix using all drug pairs. Given two drugs, $d_a$ and $d_b$, and their binary feature vectors ($u_a$ and $u_b$ $\in [0,1]^M$), their similarity was calculated according to the \textit{Jaccard} score:

\[J(u_a,u_b) = M_{11} / (M_{01} + M_{10} + M_{11}), \]
\[0 \leq J(u_a,u_b) \leq 1\]

where $M_{01}$ represents the count of positions ($i \in [1, \cdots, M]$) in $u_a$ and $u_b$ where $u_{ai} = 0$ and $u_{bi} = 1$. Similarly, $M_{10}$  represents the count of positions ($i$) in $u_a$ and $u_b$ where $u_{ai} = 1$ and $u_{bi} = 0$. Lastly, $M_{11}$ denotes the count of positions ($i$) in $u_a$ and $u_b$ where $u_{ai} = 1$ and $u_{bi} = 1$. This similarity measure is calculated for each drug pair resulting in a $N \times N$ similarity matrix.

DS2 and DS3 were generated by similar approaches. The description of the similarity matrix constructions can be found in \cite{zhang_predicting_2017} for the DS1 similarity matrices, in \cite{wan_neodti_2019} for DS2 and in \cite{gottlieb_indi_2012} for DS3.

\subsection*{Database DDI labels}
In a supervised classification setting, labels of known drug-drug interactions are required in the form of a binary matrix with the same dimensions ($N \times N$) as the input similarity matrices (Table \ref{table:labels}). For example, the labels in DS1 were provided by the \textit{TWOSIDES} database \cite{tatonetti_data-driven_2012}.

Notably, the DS3 dataset labels are split based on whether the DDIs result from a shared CYP metabolizing enzyme (\textit{CYP}) or not (\textit{NCYP}). This separation was made on the grounds that CYPs are major enzymes involved in $\sim75\%$ of the total drug metabolism. As an example, one drug would inhibit a specific CYP enzyme which also metabolizes another drug, therefore triggering a CYP-related DDI. This separation of CYP labels can affect the model training and predictability, as the positive labels are way outnumbered by the negative ones (Table \ref{table:labels}).

The known DDIs in these label matrices have the label value $1$. Label $0$, however, does not guarantee the absence of drug interactions for the given drug pair. An interaction in this case, may not have been observed yet, or may not have been included in the specific DDI database.

\begin{table}[ht]
\begin{center}
    \normalsize
\begin{tabular}{l|r|r|r|r}
Dataset & $\#$ drugs & $\#$ drug-drug pairs & $\#$ known DDIs & $\%$ known DDIs \\\hline
DS1      & 548             & 149’878                   & 48’584               & $\sim$32\%               \\
DS2      & 707             & 249’571                       & 17’206               & $\sim$7\%                \\
DS3 \textit{CYP}     & 807             & 325’221                   & 5’039                & $\sim$1.5\%              \\
DS3 \textit{NCYP}   & 807             & 325’221                   & 20’452               & $\sim$6\%               
\end{tabular}
    \end{center}
    \caption{Labels for each dataset.\label{table:labels}}
\end{table}

\subsection*{Model evaluation}

The model performance is evaluated based on standardized classification metrics. We included 1) \textit{AUC-ROC} and 2) \textit{AUC-PR}. For consistency with previous studies, denoted by \textit{AUC}, \textit{AUPR} from now on. These scores are composed by the definitions in Table \ref{table:confusionmatrix}.

\bgroup
\def\arraystretch{1.5}
\begin{table}[ht]
\begin{center}
    \normalsize
\begin{tabular}{l|l|c|c|c}
\multicolumn{2}{c}{}&\multicolumn{2}{c}{True Interactions}&\\
\cline{3-4}
\multicolumn{2}{c|}{}&Positive&Negative&\multicolumn{1}{c}{}\\
\cline{2-4}
\multirow{2}{*} \phantom{}Predicted & Positive & $TP$ & $FP$ & $\textbf{Precision} = TP / (TP + FP)$\\%[+1pt]
\cline{2-4} Interactions & Negative & $FN$ & $TN$ & \\
\cline{2-4}
%\multicolumn{1}{c}{} \\
\multicolumn{1}{c}{} & \multicolumn{1}{c}{} & \multicolumn{1}{c}{$\textbf{TPR}, \textbf{Recall} =$} & \multicolumn{1}{c}{$\textbf{FPR} =$} \\ %\multicolumn{1}{c}{$\textbf{F1} = 2TP / (2TP+FP+FN)$}\\
\multicolumn{1}{c}{} & \multicolumn{1}{c}{} & \multicolumn{1}{c}{$TP / (TP + FN)$} & \multicolumn{1}{c}{$FP / (FP + TN)$}\\
\end{tabular}
    \end{center}
    \caption{Confusion matrix.\label{table:confusionmatrix}}
\end{table}
\egroup

\textit{AUPR} is the Area Under the Precision-Recall curve and is considered the fairer measure \cite{rohani_drug-drug_2019} especially when class imbalance (i.e., unequal label distribution) is prevalent in the dataset. This is notably the case when the number of positive samples (labels with value $1$) and the number of negative samples ($0$s) are significantly imbalanced. Given the low proportions of positive samples (Table \ref{table:labels}) this is the main performance measure we focus on for the model evaluation. We furthermore computed the \textit{AUC} as standard classification metric. \textit{AUC} is the Area Under the TPR-FPR Curve, where TPR (also Recall) is the True Positive Rate and FPR is the False Positive Rate, as defined above. 
%\textit{F1} is the harmonic mean of Precision and Recall and is useful because of their trade-off (as Precision increases, usually Recall decreases, and vice versa).

\subsection*{Baseline model}
We compared our model to multiple baseline models found in the literature with special focus on \textit{NDD} \cite{rohani_drug-drug_2019} that showed high performance on DDI prediction (as reported by the authors). NDD consists of three parts: 1) In a first step, the similarity matrices are filtered based on matrix entropy scores. This aims at basing the classification only on the most informative similarity matrices and therefore excluding less informative ones using \textit{handcrafted} heuristics. 2) In a second step, the remaining similarity matrices are merged into one matrix through the \textit{SNF} method (i.e., using similarity network fusion algorithm) \cite{wang_similarity_2014}. 3) Finally, the fused matrix is used as input to a feed-forward classifier network which outputs binary DDI predictions. 

We re-implemented (to the best of our ability) \textit{NDD} using the \textit{PyTorch} deep learning library \cite{paszke_automatic_2017} for the purpose of reproducing the baseline model results. However, we were not able to reproduce the model results reported in \cite{rohani_drug-drug_2019} especially for DS2 and DS3 datasets. Therefore, we report the performance values cited by the author in their article \cite{rohani_drug-drug_2019, rohani_iscmf_2020}.

\begin{figure}[ht]
  \centering
      \includegraphics[width=0.5\textwidth]{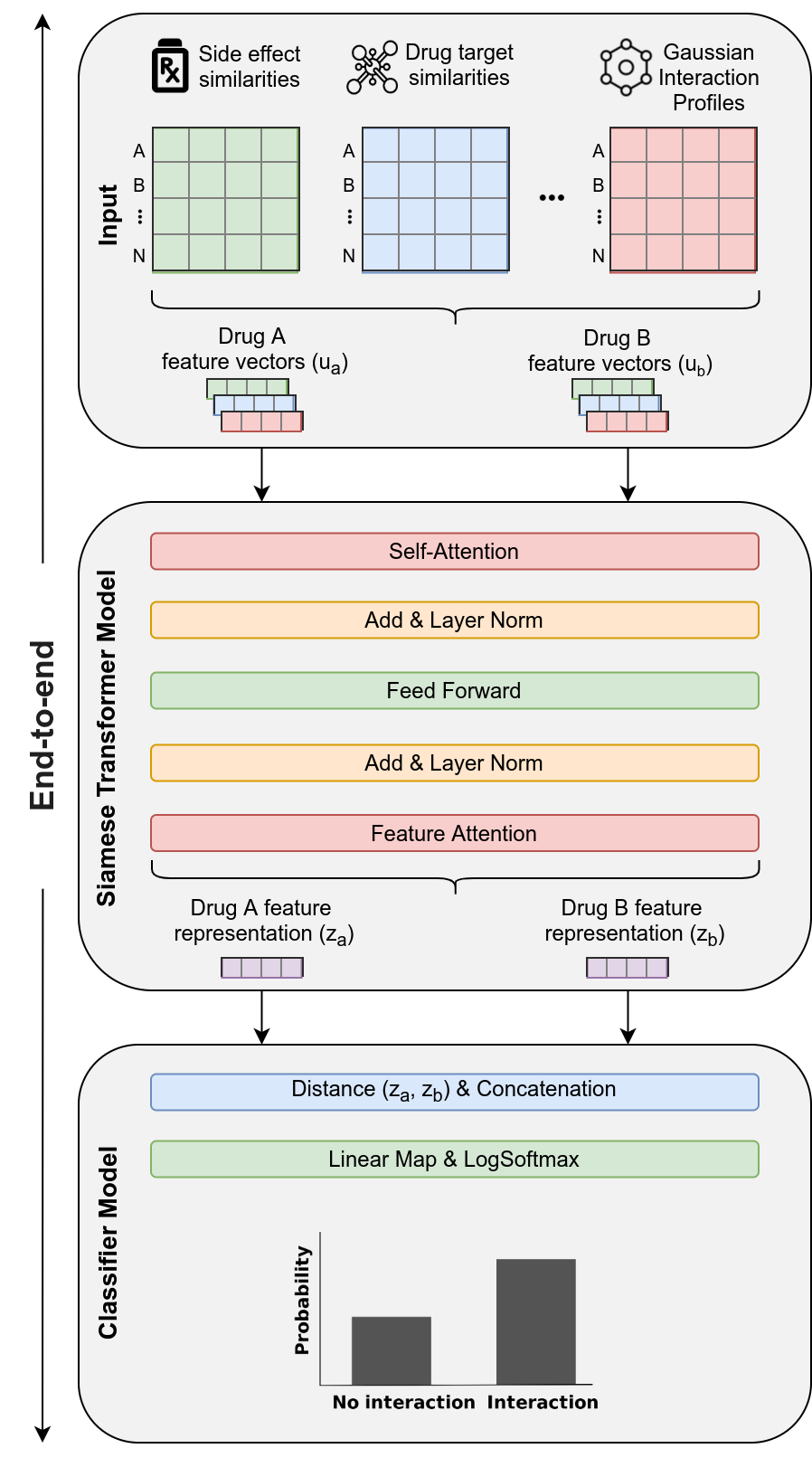}
      \caption{\textit{AttentionDDI} model architecture. 1) The sets of drug pair feature vectors $(u_a,u_b)$ from each similarity matrix are used as model input, separately for each drug. 2) A \textit{Transformer}-based Siamese encoder model generates new drug feature representation vectors for each drug. First, by applying learned weights (through \textit{Self-Attention}) to the drug feature vectors. Then, by non-linearly transforming the weighted feature vectors by a feed-forward network. Finally, a \textit{Feature Attention} pooling method aggregates the transformed feature vectors into a single feature vector representation for each drug ($z_a$ or $z_b$ respectively). 3) A separate classifier model concatenates the encoded feature vectors $z_a,z_b$ with their distance (\textit{euclidean} or \textit{cosine}). Lastly, through affine mapping of the concatenated drug pair vectors followed by \textit{Softmax} function, a drug-interaction probability distribution is generated for each drug pair.\label{fig:architecture_overview}}
\end{figure}

\subsection*{AttentionDDI: Model description}

We constructed a Siamese \textit{multi-head} \textit{self-Attention} \textit{multi-modal} neural network model (Figure \ref{fig:architecture_overview}) adapting the Transformer architecture to model our DDI problem.  
\paragraph*{Siamese model}
Our model is a Siamese neural network \cite{Chicco2021} designed to use the same model weights for processing in tandem two different input vectors. In our case the drug similarity features of each drug pair $(d_a, d_b)$ are encoded in parallel in order to learn improved latent vector representations. They are used in a later stage for computing a distance/similarity between both vectors.

% \paragraph*{Attention \& Self-Attention}
% \textit{Attention} \cite{vaswani_attention_2017} is a machine learning technique that allows a model to focus on parts of the input data or computed hidden/latent vector states as needed. That is, the model learns a vector of importance weights (i.e., normalized weights) on the input data. In our case, we are predicting DDIs based on several drug similarity matrices. \textit{Attention} can improve predictions by focusing on the most relevant similarity types. For instance, if drug indications would be more crucial for predicting DDIs compared to other similarity types (e.g., side effects, drug targets, etc.) then the \textit{Attention} mechanism would weight them more heavily.

\paragraph*{Transformer architecture}
Our model architecture adapts the \textit{Transformer} network \cite{vaswani_attention_2017} that uses \textit{multi-head} \textit{self-attention} mechanism to compute new latent vector representations from the set of input vectors while being optimized during training for our \textit{DDI} prediction problem. It consists of:

\begin{enumerate}
    \item An \textit{Encoder} model, which takes as input a set of drug similarity feature vectors and computes a new (unified) fixed-length feature vector representation. 
    \item A \textit{Classifier} model, which given the new feature vector representations, generates a probability distribution for each drug pair, indicating if this drug pair is more likely to interact or not.
\end{enumerate}

\paragraph*{Input vectors}
Our model is trained on each benchmark dataset (i.e., DS1, DS2 and DS3) separately. There are one or more similarity matrices in a given dataset and $N$ distinct number of drugs. Furthermore, there are $K = \binom{N}{2}$ drug pair combinations in every dataset. For a drug pair $(d_a,d_b)$ in a dataset $D$, the drug feature vectors $(u_a,u_b)$ each represent a set of input feature vectors extracted from corresponding similarity matrix $\{S_1, S_2, \cdots, S_T\} \in D$ (including GIP) in dataset $D$. Each set (i.e., $u_a$ and $u_b$) is used as model's input for each drug separately where $T$ feature vectors are processed. For instance, a dataset with three similarity matrices (including GIP) would have two sets of three input vectors (Figure \ref{fig:architecture_overview}) for each drug pair:
\[u_a = \{S_1^{d_a}, S_2^{d_a}, S_3^{d_a}\}\]
\[u_b = \{S_1^{d_b}, S_2^{d_b}, S_3^{d_b}\}\]

\subsection*{Encoder model}
For each drug pair $(d_a,d_b)$ the sets of drug feature vectors $(u_a,u_b)$ go through the Encoder separately, in parallel (hence, Siamese model). The Encoder consists of multiple layers. Initially, the input vectors go through a \textit{Self-Attention} layer that aims at generating improved vector encoding (i.e., new learned representation) while optimizing for the target task (i.e., classification in our setting). During this step, the drug feature vectors are weighted according to how strongly they are correlated to the other feature vectors of the same drug. Subsequently, those weighted vectors are fed into a feed-forward network in order to calculate new feature vector representations via non-linear transformation. Lastly, the encoded feature vector representations are passed through a \textit{Feature Attention} layer which aggregates the learned representations, i.e., pools across similarity type vectors. The Encoder then outputs the two separate drug representation vectors $(z_a,z_b)$ which are then fed into the Classifier model. Additionally, there are \textit{Add + Normalize} layers (i.e., residual connections and normalization) after the \textit{Self-Attention} and \textit{Feed-Forward} layers which are used for more efficient training. To summarize, the encoder consists of the following layers in this order: \textit{Self-Attention}, \textit{Add + Normalize}, \textit{Feed-Forward}, \textit{Add + Normalize}, \textit{Feature Attention}.

\subsubsection*{Self-attention layer}
We followed a multi-head self-attention approach where multiple single-head self-attention layers are used in parallel (i.e., simultaneously) to process each input vector in set $u$ (i.e., $u_a$ for drug $d_a$). The outputs from every single-head layer are concatenated and transformed to generate a fixed-length vector using an affine transformation. The single-head self-attention approach \cite{vaswani_attention_2017} performs linear transformation to every input vector using three separate matrices: (1) a queries matrix $W_{query}$, (2) keys matrix $W_{key}$, and (3) values matrix $W_{value}$. Each input $u_t$ where $t$ indexes the feature vectors in $u$ (i.e., set of input feature vectors for a given drug extracted from similarity matrices $\{S_1, S_2, \cdots, S_T\} \in D$)  is mapped using these matrices to compute three new vectors (Eq. \ref{eq:query_mat}, \ref{eq:key_mat}, and \ref{eq:value_mat})
\begin{equation}
 q_t = W_{query} u_t
\label{eq:query_mat}
\end{equation}
\begin{equation}
 k_t = W_{key} u_t
\label{eq:key_mat}
\end{equation}
\begin{equation}
 v_t = W_{value} u_t
\label{eq:value_mat}
\end{equation}
where $W_{query}$, $W_{key}$, $W_{value}$ $\in \R^{d'\times d}$, $q_t$, $k_t$, $v_t$ $\in \R^{d'}$ are query, key and value vectors, and $d'$ is the dimension of the three computed vectors respectively.
In a second step, attention scores are computed using the pairwise similarity between the query and key vectors for each input vector $u_t$ in the input set $u$. The similarity is defined by computing a scaled dot-product between the pairwise vectors. For each input vector, we compute attention scores $\alpha_{tl}$ representing the similarity between $q_t$ and vectors $k_l$ $\forall l \in [1, \dots , T]$ where $T$ representing the number of vectors in the input set $u$ (Eq. \ref{eq:attn_shl_alpha}, \ref{eq:attn_shl_score}) and then normalized using $softmax$ function. Then a weighted sum using the attention scores $\alpha_{tl}$ and value vectors $v_l$ $\forall l \in [1, \dots , T]$ is performed (Eq. \ref{eq:attn_shl_wsum}) to generate a new vector representation $r_t \in \R^{d'}$ for the input vector $u_t$. This process is applied to every input vector in the input set $u$ to obtain a new set of input vectors $\underline{R} = \{r_1,r_2, \cdots, r_{T}\}$.
\begin{equation}
\alpha_{tl} = \frac{\exp{(score(q_t, k_l))}}{\sum_{l=1}^{T}\exp{(score(q_t,k_l))}}
\label{eq:attn_shl_alpha}
\end{equation}
\begin{equation}
score(q_t, k_l) = \frac{{q_t}^\top k_l}{\sqrt{d'}}
\label{eq:attn_shl_score}
\end{equation}
\begin{equation}
r_t = \sum_{l=1}^T \alpha_{tl}v_l
\label{eq:attn_shl_wsum}
\end{equation}
In a multi-head setting with $H$ number of heads, the queries, keys and values matrices will be indexed by superscript $h$ (i.e., 
$W^h_{query}$, $W^h_{key}$, $W^h_{value}$ $\in \R^{d'\times d}$) and applied separately to generate a new vector representation $r^h_t$ for every single-head self-attention layer. The output from each single-head layer is concatenated into one vector $r^{concat}_t = concat(r^1_t, r^2_t, \cdots, r^H_t)$ where $r^{concat}_t \in \R^{d'H}$ and then transformed using affine transformation (Eq. \ref{eq:attn_mh}) such that $W_{unify} \in \R^{d'\times d'H}$ and $b_{unify} \in \R^{d'}$. This process is applied to each position in the set $\underline{R}$ to generate a new set of vectors  $\underline{\tilde{R}} = \{\tilde{r}_1,\tilde{r}_2, \cdots, \tilde{r}_T\}$.
\begin{equation}
\tilde{r_t} = W_{unify} r^{concat}_t + b_{unify}
\label{eq:attn_mh}
\end{equation}

\subsubsection*{Layer Normalization \& Residual Connections}
We used residual/skip connections \cite{He2016} in order to improve the gradient flow in layers during training. This is done by summing both the newly computed output of the current layer with the output from the previous layer. In our setting, a first residual connection sums the output of the self-attention layer $\tilde{r}_t$ and the input vector $u_t$ for every feature vector in the input set $u$. We will refer to the summed output by $\tilde{r_t}$ for simplicity. \newline
Layer normalization \cite{Ba2016} was used in two occasions; after the self-attention layer and the feed-forward network layer with the goal to ameliorate the "covariate-shift" problem by re-standardizing the computed vector representations (i.e., using the mean and variance across the features/embedding dimension $d'$). Given a computed vector $\tilde{r_t}$, $LayerNorm$ function will standardize the input vector using the mean $\mu_t$ and variance $\sigma_t^2$ along the features dimension $d'$ and apply a scaling $\gamma$ and shifting step $\beta$ (Eq. \ref{eq:layernorm_func}). $\gamma$ and $\beta$ are learnable parameters and $\epsilon$ is small number added for numerical stability.\newline

\begin{equation}
\mu_t = \frac{1}{d'}\sum_{j=1}^{d'}\tilde{r}_{tj}
\label{eq:layernorm_mean}
\end{equation}
\begin{equation}
\sigma^2_t = \frac{1}{d'}\sum_{j=1}^{d'}(\tilde{r}_{tj} - \mu_t)^2
\label{eq:layernorm_variance}
\end{equation}
\begin{equation}
LayerNorm(\tilde{r}_t) = \gamma \times \frac{\tilde{r}_t - \mu_t}{\sqrt{\sigma^2_t + \epsilon}} + \beta
\label{eq:layernorm_func}
\end{equation}

\subsubsection*{FeedForward Layer}
After a layer normalization step, a feed-forward network consisting of two affine transformation matrices and non-linear activation function is used to further compute/embed the learned vector representations from previous layers. The first transformation (Eq. \ref{eq:feedforward}) uses $W_{MLP1} \in \R^{\xi d' \times d'}$ and $b_{MLP1} \in \R^{\xi d'}$ to transform input $\tilde{r_t}$ to new vector $\in \R^{\xi d'}$ where $\xi \in \mathbb{N}$ is multiplicative factor. A non-linear function such as $ReLU(z)=max(0,z)$ is applied followed by another affine transformation using $W_{MLP2} \in \R^{d'\times \xi d'}$ and $b_{MLP2} \in R^{d'}$ to obtain vector $g_t \in \R^{d'}$. A layer normalization (Eq. \ref{eq:layernorm_2}) is applied to obtain $\tilde{g}_t \in \R^{d'}$. \newline
\begin{equation}
g_t = W_{MLP2} ReLU(W_{MLP1} \tilde{r_t} + b_{MLP1}) + b_{MLP2}
\label{eq:feedforward}
\end{equation}

\begin{equation}
\tilde{g}_t = LayerNorm(g_t)
\label{eq:layernorm_2}
\end{equation}
These transformations are applied to each vector in set $\underline{\tilde{R}}$ to obtain new set $\underline{\tilde{G}} = \{\tilde{g}_1, \tilde{g}_2, \cdots, \tilde{g}_T\}$. At this point, the \emph{encoder} block operations are done and multiple encoder blocks can be stacked in series for $E$ number of times. In our experiments, $E$ was a hyperparameter that was empirically determined using a validation set (as the case of the number of attention heads $H$ used in self-attention layer).

\subsubsection*{Feature Attention Layer}
The feature attention layer is parameterized by a \emph{global} context vector $c$  with learnable parameters optimized during the training. For a set of input vectors $\underline{\tilde{G}} = \{\tilde{g}_1, \tilde{g}_2, \cdots, \tilde{g}_T\}$ (computed in the layer before), attention scores $\psi_t \forall t \in [1, \cdots, T]$ are calculated using the pairwise similarity between the context vector $c \in \R^{d'}$ and the set $\underline{\tilde{G}}$ (Eq. \ref{eq:attn_val_poslayer}, \ref{eq:attn_score_poslayer}). These scores are normalized and used to compute weighted sum of the $\{\tilde{g}_1, \tilde{g}_2, \cdots, \tilde{g}_T\}$ vectors to generate a new \textit{unified} vector representation $z \in \R^{d'}$ that is further passed to the classifier layer.
\begin{equation}
\psi_{t} = \frac{\exp{(score(c, \tilde{g}_t))}}{\sum_{j=1}^{T}\exp{(score(c,\tilde{g}_j))}}
\label{eq:attn_val_poslayer}
\end{equation}
\begin{equation}
score(c, \tilde{g}_t) = \frac{{c}^\top \tilde{g}_t}{\sqrt{d'}}
\label{eq:attn_score_poslayer}
\end{equation}
\begin{equation}
z = \sum_{t=1}^T \psi_{t}\tilde{g}_t
\label{eq:attn_wsum_poslayer}
\end{equation}

\paragraph*{Classifier layer}
The classifier layer calculates a distance (\textit{euclidean} or \textit{cosine}) between the computed representation vectors $(z_a,z_b)$ and then concatenates them with that distance. Subsequently, through an affine transformation, the concatenated feature vector is mapped to the size of the output classes (i.e., presence or absence of interaction). Finally, a \textit{softmax} function is applied to output the predicted probability distribution over those two classes.

\subsection*{Objective Function}
We defined the total loss for an $i$-th drug pair by a \textit{linear} combination of the negative log-likelihood loss ($L^C$) and the contrastive loss ($L^{Dist}$). The contribution of each loss function is determined by a hyperparameter $\gamma \in  (0,1)$. Additionally,  a weight regularization term (i.e., $l_2$-norm regularization) applied to the model parameters represented by $\bm{\theta}$ is added to the objective function.

% $\binom{N}{2}$

\begin{equation}
    L^{Total} = \gamma L^C + (1-\gamma) L^{Dist}\ + \frac{\lambda}{2}||\mathbb{\bm{\theta}}||_{2}^{2}
\label{eq:total_loss}
\end{equation}

where

\begin{equation}
    l^{C}_{(i)} = - [y_{(i)} log \hat{y}_{(i)} + (1-y_{(i)})log(1- \hat{y}_{(i)})], y_i \in \{0,1\}
\end{equation}
\begin{equation}
    L^{C} = \frac{1}{K}\sum_{i=1}^{K} l^{C}_{(i)}
\end{equation}

and 

\begin{equation}
    l^{Dist}_{(i)} = \begin{cases}
y_i = 1 & \frac{1}{2} {Dist}^2_{(i)}\\
y_i = 0 & \frac{1}{2} max((\mu - {Dist}_{(i)})^2, 0)
\end{cases}
\end{equation}

\begin{equation}
    L^{Dist} = \frac{1}{K}\sum_{i=1}^{K} l^{Dist}_{(i)}
\end{equation}

$Dist_{(i)}$ represents the computed distance between the encoded vector representations $z_a$ and $z_b$ of $i^{th}$ drug pair, which can be \textit{euclidean} or \textit{cosine} distance. Additionally, $\mu$ is a contrastive loss \textit{margin} hyperparameter.

The training is done using mini-batches where computing the loss function and updating the parameters/weight occur after processing each mini-batch of the training set.

\subsection*{Training workflow}
For training, we utilized a $10$-fold stratified cross-validation strategy with $10\%$ dedicated for validation set along with the hyperparameters defined in Table \ref{table:hyperparam}. During training, examples were weighted inversely proportional to class/outcome frequencies in the training data. Model performance was evaluated using area under the receiver operating characteristic curve (AUC), and area under the precision recall curve (AUPR). During training of the models, the epoch in which the model achieved the best AUPR on the validation set was recorded, and model state as it was trained up to that epoch was saved. This best model, as determined by the validation set, was then tested on the test split.

\begin{table}[htbp]
\begin{center}
    \normalsize
\begin{tabular}{l|r|r|r|r}
              & \textbf{DS1}      & \textbf{DS2}      & \textbf{DS3 CYP}  & \textbf{DS3 NCYP} \\ \hline
\# attention heads ($H$)  & $2$        & $2$        & $4$        & $2$        \\
\# transformer units ($E$) & $1$        & $1$        & $1$        & $1$        \\
Dropout              & $0.3$      & $0.3$      & $0.45$     & $0.3$      \\
MLP embed factor ($\xi$)    & $2$        & $2$        & $2$        & $2$        \\
Pooling mode         & attn   & attn   & attn   & attn   \\ Distance             & cosine & cosine & cosine & cosine \\
Weight decay         & $1^{-6}$ & $1^{-6}$ & $1^{-8}$ & $1^{-6}$ \\
Batch size           & $1000$     & $1000$     & $400$      & $1000$     \\
\# epochs            & $100$      & $100$      & $200$      & $100$      \\
$\gamma$                    & $0.05$     & $0.05$     & $0.05$     & $0.05$     \\
$\mu$                    & $1$        & $1$        & $1$        & $1$       
\end{tabular}
    \end{center}
    \caption{Training hyperparameters.\label{table:hyperparam}}
\end{table}

\clearpage
\section*{Results}

\paragraph*{Model evaluation results}
We compared our model \textit{AttentionDDI} against state-of-the-art models, as shown in Table \ref{table:model_scores_DS_all}. Our model overall achieves similar or better prediction performance when tested on four distinct benchmark datasets.

For DS1, our model achieves an AUPR score of $0.924$, outperforming the baseline NDD model (AUPR $0.922$). The best performing model for DS1 is the Classifier ensemble model (AUPR $0.928$). For DS2 our model outperforms all models with an AUPR score of $0.904$, with NDD coming second with an AUPR score of $0.89$. For DS3 with the CYP labels, our model achieves the second best AUPR score of $0.775$, surpassed by the baseline model (AUPR $0.830$). We would like to note that most models perform poorly (AUPR $< 0.5$) on this dataset. Finally, for DS3 with NCYP labels our model with AUPR score of $0.890$ vastly outperforms most models, except for the NDD model (AUPR $0.947$).

\begin{table}[htbp]
\begin{center}
    \normalsize
\begin{tabular}{l|c|c|c|c|c|c|c|c}
\textbf{Model}                       & \multicolumn{2}{c|}{\textbf{DS1}} & \multicolumn{2}{c|}{\textbf{DS2}} & \multicolumn{2}{c|}{\textbf{DS3 (CYP)}} & \multicolumn{2}{c}{\textbf{DS3 (NCYP)}} \\ \hline
\textbf{Score}                       & \textbf{AUC}        & \textbf{AUPR}       & \textbf{AUC}        & \textbf{AUPR}       & \textbf{AUC}           & \textbf{AUPR}          & \textbf{AUC}            & \textbf{AUPR}          \\ \hline
\textit{\textbf{AttentionDDI}} (our model)                & \textbf{0.954}      & \textbf{0.924}     & \textbf{0.986}      & \textbf{0.904}      &\textbf{0.989}         & \textbf{0.775}         & \textbf{0.986}          & \textbf{0.890}         \\
\textit{\textbf{NDD}}*                        & 0.954      & 0.922      &\textbf{0.994}    & 0.890       &\textbf{0.994}        &\textbf{0.830}        &\textbf{0.992}         &\textbf{0.947}        \\
\textit{Classifier ensemble}*        & 0.956      &\textbf{0.928}     & 0.936      & 0.487      & 0.990          & 0.541         & 0.986          & 0.756         \\
\textit{Weighted average ensemble}*  & 0.948      & 0.919      & 0.646      & 0.440       & 0.695         & 0.484         & 0.974          & 0.599         \\
\textit{RF}*                         & 0.830       & 0.693      & 0.982      & 0.812      & 0.737         & 0.092         & 0.889          & 0.167         \\
\textit{LR}*                         & 0.941      & 0.905      & 0.911      & 0.251      & 0.977         & 0.487         & 0.916          & 0.472         \\
\textit{Adaptive boosting}*          & 0.722      & 0.587      & 0.904      & 0.185      & 0.830          & 0.143         & 0.709          & 0.150          \\
\textit{LDA}*                        & 0.935      & 0.898      & 0.894      & 0.215      & 0.953         & 0.327         & 0.889          & 0.414         \\
\textit{QDA}*                        & 0.857      & 0.802      & 0.926      & 0.466      & 0.709         & 0.317         & 0.536          & 0.260          \\
\textit{KNN}*                        & 0.730       & 0.134      & 0.927      & 0.785      & 0.590          & 0.064         & 0.603          & 0.235         \\
\textit{\textbf{ISCMF}}\textdagger                     & 0.899      & 0.864      &    -        &      -      & 0.898         & 0.767         & 0.898          & 0.792         \\
\textit{Classifier ensemble}\textdagger       &\textbf{0.957}      & 0.807      &    -        &        -    & 0.990          & 0.541         & 0.986          & 0.756         \\
\textit{Weighted average ensemble}\textdagger & 0.951      & 0.795      &     -       &      -      & 0.695         & 0.484         & 0.974          & 0.599         \\
\textit{Matrix perturbation}\textdagger       & 0.948      & 0.782      &    -        &      -      &        -       &       -        &          -      &        -       \\
\textit{Neighbor Recommender}\textdagger      &          -  &    -        &      -      &    -        & 0.953         & 0.126         & 0.904          & 0.295         \\
\textit{Label Propagation}\textdagger         &    -        &        -    &      -      &    -        & 0.952         & 0.126         &          -      &        -       \\
\textit{Random walk}\textdagger               &      -      &    -        &      -      &    -        &      -         &         -      & 0.895          & 0.181        
\end{tabular}
    \end{center}
    \caption[]
    {\tabular[t]{@{}l@{}}Model evaluation scores for all datasets.\\
    *: scores from \cite{rohani_drug-drug_2019}, \textdagger: scores from \cite{rohani_iscmf_2020}\endtabular
    \label{table:model_scores_DS_all}}
\end{table}

\paragraph*{Attention weights}
\newcommand{\attnweightwidth}{0.666\textwidth}

\begin{figure}[htbp]
  \centering
      \includegraphics[width=\attnweightwidth]{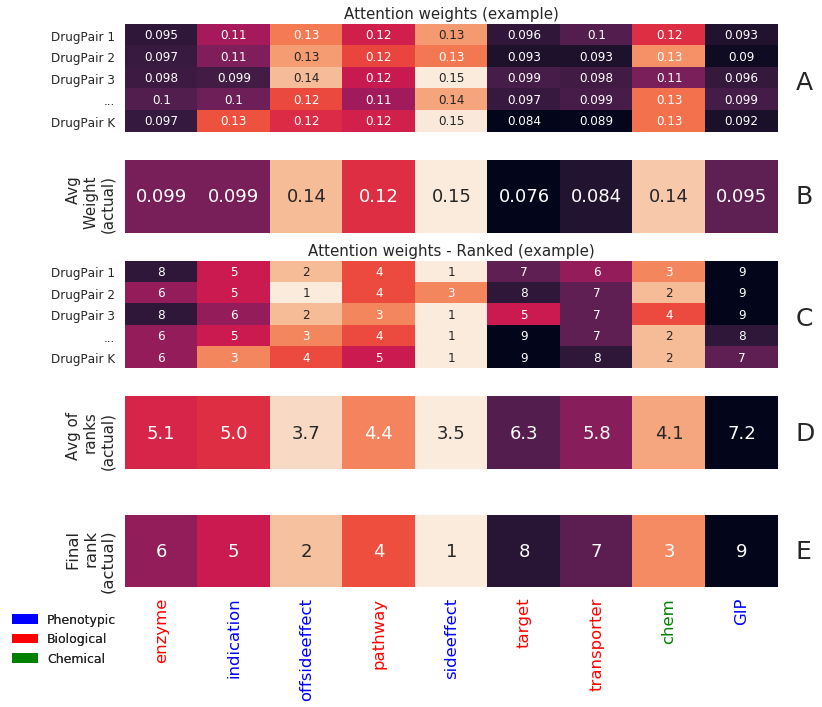}
      \caption{
     \textit{Feature Attention} weights for DS1. \textbf{A}: \textit{Example} weights for each drug pair. Each row represents the weights of the \textit{Feature Attention} layer for each drug pair $(a,b)$ averaged. Since there are $K$ drug pairs in total and $9$ drug similarity matrices (columns) this is a $K \times 9$ matrix. The default value is $0.11 (1/9)$ and each row sums to $1$. \textbf{B}: Average weights per column (similarity type) for the \textit{actual} weights in the $K \times 9$ matrix from \textbf{A}. \textbf{C}: The absolute \textit{example} weights for each row in \textbf{A} are ranked from highest ($1$) to lowest ($9$). \textbf{D}: Average ranks per column (similarity type) for the \textit{actual} ranks in the $K \times 9$ matrix from \textbf{C}. \textbf{E}: The actual average values from \textbf{D} are ranked again to obtain the final similarity type rank from most important ($1$) to least important ($9$). Additionally, the similarity type labels were manually colored to indicate the level of information (phenotypic, biological, chemical).
      \label{fig:attn_weights_DS1}}
\end{figure}

\begin{figure}[htbp]
  \centering
      \includegraphics[width=\attnweightwidth]{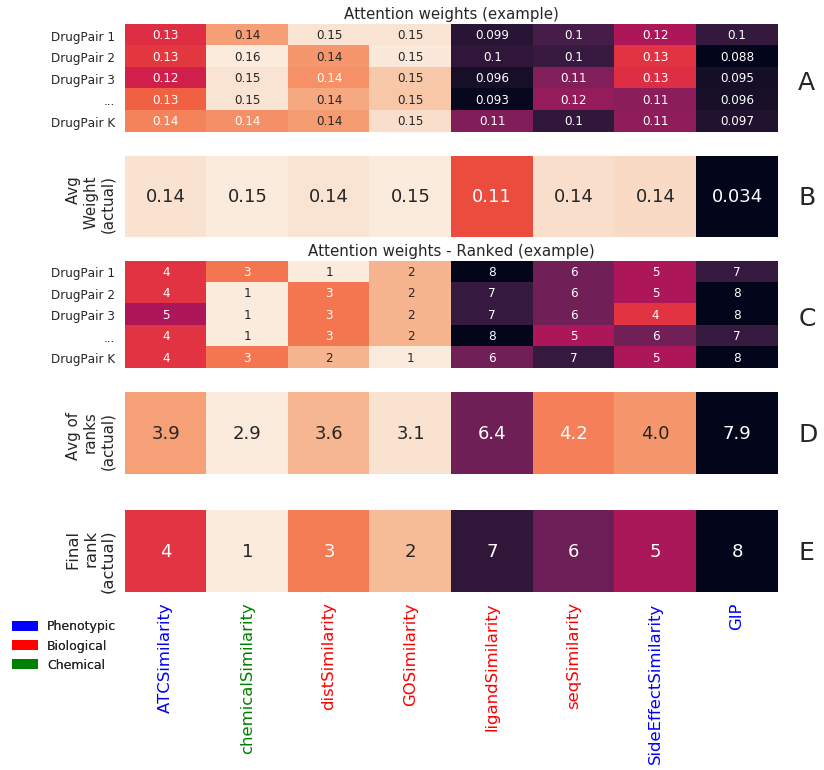}
      \caption{
     \textit{Feature Attention} weights for DS3 with the CYP labels.
      \label{fig:attn_weights_DS3_CYP}}
\end{figure}

\begin{figure}[htbp]
  \centering
      \includegraphics[width=\attnweightwidth]{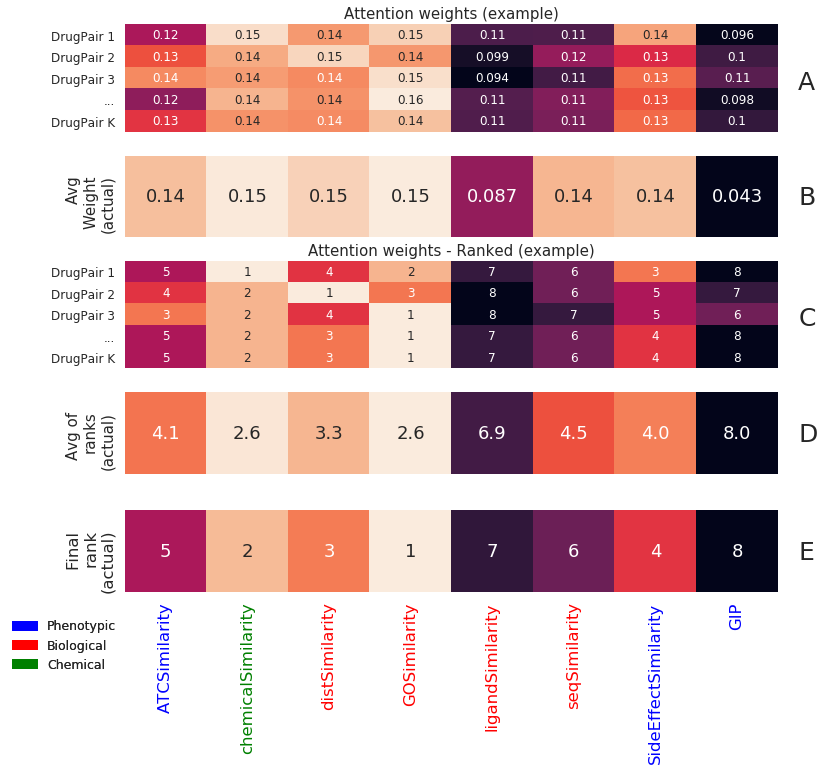}
      \caption{
     \textit{Feature Attention} weights for DS3 with the NCYP labels.
      \label{fig:attn_weights_DS3_NCYP}}
\end{figure}

Our model offers model explainability through the \textit{Feature Attention} layer (Figure \ref{fig:architecture_overview}). This layer determines the contribution (weight) of the similarity matrices to each of the encoded vector representations ($z_a, z_b$). Those weights are illustrated in Figure \ref{fig:attn_weights_DS1} for DS1 and in Figures \ref{fig:attn_weights_DS3_CYP} and \ref{fig:attn_weights_DS3_NCYP} for the labels \textit{CYP} and \textit{NCYP} of DS3, accordingly. 

For DS1, the default similarity weight is $0.11$. The top three ranked similarity matrices that were also weighted more heavily by the \textit{Feature Attention} layer for DDI predictions were \textit{sideeffect}, \textit{offsideeffect} and \textit{chem}, with average weights of $0.15$, $0.14$ and $0.14$ while most other matrices were pushed below $0.1$. This shows that the main contributors for DDI predictions contain high-level phenotypic information (drug side effects) while drug chemical structures also play a major role. 

In the DS3 dataset, for both the CYP and NCYP labels, the top three ranked similarity matrices were \textit{chemicalSimilarity}, (PPI) \textit{distSimilarity} and \textit{GOSimilarity}, though with different ranking orders. According to the average weights, for this dataset the weights were more evenly distributed with the top three getting a weight of $0.15$, as the default weight is $0.125$. All three similarity groups, containing phenotypic, biological and chemical information have a similar contribution for the DDI predictions in DS3. Notably, \textit{ligandSimilarity} and \textit{GIP} were weighted much lower than the other similarity matrices, possibly due to this type of information not leading to good DDI predictions.

\paragraph*{Case Studies}
To further test the efficiency of our model, we investigated the top predictions of our model through an external drug interaction database, DrugBank \cite{wishart_drugbank_2018}, which contains DDIs extracted from drug labels and scientific publications. We focused on the DS1 dataset, which links drug similarities to external drug IDs and therefore can be used for external validation. From DS1, we selected the top $20$ novel predictions ("false positives" according to the DS1 labels) with the highest interaction probabilities from our model, \textit{AttentionDDI}. In Table \ref{table:case_studies} we list those drug pairs along with the interaction information from DrugBank. We found that $60\%$ of those top predictions were externally confirmed as known drug pair interactions.

% \begin{table}[ht]
% \begin{center}
    % \normalsize
    % \begin{tabular}{r|l|l|l|l|p{6cm}}
\begin{table}[ht]
\centering
\resizebox{\textwidth}{!}{\begin{tabular}{rllllp{8cm}}
\textbf{Rank} & \textbf{ID A}    & \textbf{ID B}    & \textbf{Drug A}              & \textbf{Drug B}                & \textbf{Interaction}                                                                                                   \\ \hline\noalign{\vskip 0.1cm}
1    & DB01194 & DB00273 & Brinzolamide        & Topiramate            & The risk or severity of adverse effects can be increased when Topiramate is combined with Brinzolamide.       \\ \hline\noalign{\vskip 0.1cm}
2    & DB01589 & DB00678 & Quazepam            & Losartan              & The metabolism of Quazepam can be decreased when combined with Losartan.                                      \\ \hline\noalign{\vskip 0.1cm}
3    & DB01212 & DB00417 & Ceftriaxone         & PenicillinV           & No interactions                                                                                               \\ \hline\noalign{\vskip 0.1cm}
4    & DB01586 & DB00951 & Ursodeoxycholicacid & Isoniazid             & No interactions                                                                                               \\ \hline\noalign{\vskip 0.1cm}
5    & DB01337 & DB00565 & Pancuronium         & Cisatracurium & Pancuronium may increase the central nervous system depressant (CNS depressant) activities of Cisatracurium.  \\ \hline\noalign{\vskip 0.1cm}
6    & DB00351 & DB00484 & Megestrolacetate    & Brimonidine           & No interactions                                                                                               \\ \hline\noalign{\vskip 0.1cm}
7    & DB00530 & DB00445 & Erlotinib           & Epirubicin            & No interactions                                                                                               \\ \hline\noalign{\vskip 0.1cm}
8    & DB00458 & DB00659 & Imipramine          & Acamprosate           & No interactions                                                                                               \\ \hline\noalign{\vskip 0.1cm}
9    & DB01586 & DB00319 & Ursodeoxycholicacid & Piperacillin          & No interactions                                                                                               \\ \hline\noalign{\vskip 0.1cm}
10   & DB00443 & DB00333 & Betamethasone       & Methadone             & The metabolism of Methadone can be increased when combined with Betamethasone.                                \\ \hline\noalign{\vskip 0.1cm}
11   & DB00458 & DB00321 & Imipramine          & Amitriptyline         & The metabolism of Amitriptyline can be decreased when combined with Imipramine.                               \\ \hline\noalign{\vskip 0.1cm}
12   & DB00790 & DB00584 & Perindopril         & Enalapril             & The risk or severity of angioedema can be increased when Enalapril is combined with Perindopril.              \\ \hline\noalign{\vskip 0.1cm}
13   & DB01059 & DB00448 & Norfloxacin         & Lansoprazole          & No interactions                                                                                               \\ \hline\noalign{\vskip 0.1cm}
14   & DB00571 & DB01203 & Propranolol         & Nadolol               & Propranolol may increase the arrhythmogenic activities of Nadolol.                                            \\ \hline\noalign{\vskip 0.1cm}
15   & DB00975 & DB00627 & Dipyridamole        & Niacin                & No interactions                                                                                               \\ \hline\noalign{\vskip 0.1cm}
16   & DB00967 & DB01173 & Desloratadine       & Orphenadrine          & Desloratadine may increase the central nervous system depressant (CNS depressant) activities of Orphenadrine. \\ \hline\noalign{\vskip 0.1cm}
17   & DB00222 & DB00328 & Glimepiride         & Indomethacin          & The protein binding of Glimepiride can be decreased when combined with Indomethacin.                          \\ \hline\noalign{\vskip 0.1cm}
18   & DB00193 & DB01183 & Tramadol            & Naloxone              & The metabolism of Naloxone can be decreased when combined with Tramadol.                                      \\ \hline\noalign{\vskip 0.1cm}
19   & DB00904 & DB00918 & Ondansetron         & Almotriptan           & The risk or severity of adverse effects can be increased when Ondansetron is combined with Almotriptan.       \\ \hline\noalign{\vskip 0.1cm}
20   & DB00423 & DB00794 & Methocarbamol       & Primidone             & The risk or severity of adverse effects can be increased when Methocarbamol is combined with Primidone.      
% \end{tabular}
    % \end{center}
    % \caption{Case studies for the top predictions in DS1.\label{table:case_studies}}
% \end{table}
\end{tabular}}
    \caption{Case studies for the top predictions in DS1. Interaction information from the DrugBank database.\label{table:case_studies}}
\end{table}

\section*{Discussion}
\subsection*{End-to-end solution}
In this work, we presented an end-to-end architecture that utilizes attention mechanism to train a DDI prediction model. When looking at the DDI models reported in the literature, most of them consist of separate steps for model training. For example, the two competing baseline models (NDD and ISCMF) consist of multiple cascaded steps such as 1) matrix selection/filtering, 2) matrix fusion, and 3) classification that are optimized separately during model training. Preferably, the matrix selection would be informed by the classification goal which would optimize this selection. However, the first two steps (matrix filtering and fusion) are independent from classification and therefore not informed by the model training task. In contrast, our model uses a holistic approach in which all computational steps are connected and optimized while minimizing the loss function of our classifier.  Consequently, our model is able to optimize the input information for DDI predictions at every computational step.

\subsection*{Explainability}
Along with DDI predictions, our model makes it possible to gain additional information, which is the learned focus of the model on the input features. All similarity matrices are utilized as input data without being filtered. Hence, during training, the model learns which input information is more relevant for the classification task and weighs it accordingly. This is  advantageous because the less relevant information is not completely discarded (as in the baseline model), but still taken into account as it may still provide some useful input for improved predictions. 

Moreover, when looking at the relative importance (i.e., rank) of the attention weights, the phenotypic information such as drug side effect similarities were ranked higher than the lower level information (biological) in DS1 (Figure \ref{fig:attn_weights_DS1}). This agrees with the conclusion in \cite{zhang_label_2015}, as phenotypic information is considered more informative for DDI predictions while biological and chemical information might provide less translational power. In DS3 for both the CYP, as well as the NCYP labels, the similarity matrices were more evenly weighted (Figures \ref{fig:attn_weights_DS3_CYP} and \ref{fig:attn_weights_DS3_NCYP}) without showing a clear dominant information modality.
Hence, we have shown the potential of applying \textit{Attention} based models on multi-modal biological dataset (i.e., on drug similarity features like side effects, drug targets, chemical structure, etc.) by highlighting the input that is most relevant for DDI prediction.

\subsection*{Weighing the loss functions}
Our model's loss function was defined by a linear combination of two loss functions: (1) the negative log-likelihood loss (NLL) and (2) the contrastive loss (equation \ref{eq:total_loss}). The contribution of the NLL loss was included as a standardized loss used in classification tasks. On the other hand, the contrastive loss focuses on minimizing the intra-class distances (among positive or negative samples) and maximize the inter-class distances (between positive and negative samples).

In our experiments, the importance of contrastive loss over the NLL loss became evident especially for DS3 datasets. For DS1 and DS2, a uniform weight between both losses would result in similar performance as reported in the manuscript. However, biasing the weight towards contrastive loss (i.e., $95\%$ more importance for contrastive loss) helped increasing the performance by approximately 1 to 2 $\%$ for DS1 or DS2.
However, for the DS3 dataset, weighing heavily (i.e., $95\%$) the contrastive loss was they key factor for achieving the high performance reported in the results section. This could be an indication that the positive and negative samples (that lead to drug interactions or not) are in close distance to each other and not well separated. In such a case, the contrastive loss would assist in better separating those samples and hence improve model performance. This was pronounced in the case of the DS3 dataset, where the proportions of positive samples are low ($\sim1.5\%$ for CYP, $\sim6\%$ for NCYP). Accordingly, the contrastive loss helped in the imbalanced label setting where the separability between classes is harder due to the low number of positive samples.

\subsection*{Conclusions}
DDIs have important implications on patient treatment and safety. Due to the large number of possible drug pair combinations, many possible DDIs remain to be discovered. Thus, DDI prediction methods, and particularly computational methods, can aid in the accelerated discovery of additional interactions. These results are valuable for healthcare professionals that aim at finding the most effective treatment combinations while attempting to minimize unintended drug side effects.

In conclusion,, we present a novel DDI prediction solution which employs \textit{Attention}, a mechanism that has successfully advanced model performance in other domains (such as \textit{NLP}). We demonstrated that \textit{Attention} based models can be  successfully adapted to multi-modal biological data in the DDI domain with increased DDI prediction performance over various benchmark datasets and enhanced model explainability. 

%%%%%%%%%%%%%%%%%%%%%%%%%%%%%%%%%%%%%%%%%%%%%%
%%                                          %%
%% Backmatter begins here                   %%
%%                                          %%
%%%%%%%%%%%%%%%%%%%%%%%%%%%%%%%%%%%%%%%%%%%%%%

\subsection*{Availability of data and materials}
The preprocessing scripts and the models’ implementation (training and testing) workflow is made publicly available at \url{https://github.com/uzh-dqbm-cmi/side-effects/tree/attn_siamese_scheduler_clean}

\subsection*{Competing interests}
    The authors declare that they have no competing interests.

\subsection*{Acknowledgements}
NA.

\subsection*{Author's contributions}
KS and AA worked on the development of processing and analysis workflow, algorithms and models implementation. KS, AA, and NPG analyzed and interpreted the data. KS drafted the manuscript. NPG, AA, and MK supervised and edited the manuscript. All authors approved the final article.

\section*{Acronyms}

%\begin{center}
\begin{tabular}{ l l }
 \textbf{ATC} & Anatomical Therapeutic Chemical \\
 \textbf{CYP} & Cytochrome P450 \\
 \textbf{DDI} & Drug-Drug Interaction \\ 
 \textbf{GIP} & Gaussian Interaction Profile \\
 \textbf{GO} & Gene Ontology \\
 \textbf{NLP} & Natural Language Processing \\
 \textbf{PPI} & Protein-Protein Interaction \\
 \textbf{SNF} & Similarity Network Fusion \\
 \\
 \textbf{TP} & True Positive \\
 \textbf{TN} & True Negative \\
 \textbf{FP} & False Positive \\
 \textbf{FN} & False Negative \\
 \\
 \textbf{TPR} & True Positive Rate \\
 \textbf{FPR} & False Positive Rate \\
 \\
 \textbf{AUC} & Area Under Curve \\
 \textbf{NLL} & Negative Log-Likelihood \\
 \textbf{ROC} & Receiver Operating Characteristic \\
 \textbf{PR} & Precision-Recall
 
\end{tabular}
%\end{center}

%%  After submission of the .TEX file,                     %%
%%  you will be prompted to submit your .BBL file.         %%

% if your bibliography is in bibtex format, use those commands:
\bibliographystyle{bmc-mathphys} % Style BST file (bmc-mathphys, vancouver, spbasic).
\bibliography{article.bib}      % Bibliography file (usually '*.bib' )


\begin{thebibliography}{18}
\ifx \bisbn   \undefined \def \bisbn  #1{ISBN #1}\fi
\ifx \binits  \undefined \def \binits#1{#1}\fi
\ifx \bauthor  \undefined \def \bauthor#1{#1}\fi
\ifx \batitle  \undefined \def \batitle#1{#1}\fi
\ifx \bjtitle  \undefined \def \bjtitle#1{#1}\fi
\ifx \bvolume  \undefined \def \bvolume#1{\textbf{#1}}\fi
\ifx \byear  \undefined \def \byear#1{#1}\fi
\ifx \bissue  \undefined \def \bissue#1{#1}\fi
\ifx \bfpage  \undefined \def \bfpage#1{#1}\fi
\ifx \blpage  \undefined \def \blpage #1{#1}\fi
\ifx \burl  \undefined \def \burl#1{\textsf{#1}}\fi
\ifx \doiurl  \undefined \def \doiurl#1{\textsf{#1}}\fi
\ifx \betal  \undefined \def \betal{\textit{et al.}}\fi
\ifx \binstitute  \undefined \def \binstitute#1{#1}\fi
\ifx \binstitutionaled  \undefined \def \binstitutionaled#1{#1}\fi
\ifx \bctitle  \undefined \def \bctitle#1{#1}\fi
\ifx \beditor  \undefined \def \beditor#1{#1}\fi
\ifx \bpublisher  \undefined \def \bpublisher#1{#1}\fi
\ifx \bbtitle  \undefined \def \bbtitle#1{#1}\fi
\ifx \bedition  \undefined \def \bedition#1{#1}\fi
\ifx \bseriesno  \undefined \def \bseriesno#1{#1}\fi
\ifx \blocation  \undefined \def \blocation#1{#1}\fi
\ifx \bsertitle  \undefined \def \bsertitle#1{#1}\fi
\ifx \bsnm \undefined \def \bsnm#1{#1}\fi
\ifx \bsuffix \undefined \def \bsuffix#1{#1}\fi
\ifx \bparticle \undefined \def \bparticle#1{#1}\fi
\ifx \barticle \undefined \def \barticle#1{#1}\fi
\ifx \bconfdate \undefined \def \bconfdate #1{#1}\fi
\ifx \botherref \undefined \def \botherref #1{#1}\fi
\ifx \url \undefined \def \url#1{\textsf{#1}}\fi
\ifx \bchapter \undefined \def \bchapter#1{#1}\fi
\ifx \bbook \undefined \def \bbook#1{#1}\fi
\ifx \bcomment \undefined \def \bcomment#1{#1}\fi
\ifx \oauthor \undefined \def \oauthor#1{#1}\fi
\ifx \citeauthoryear \undefined \def \citeauthoryear#1{#1}\fi
\ifx \endbibitem  \undefined \def \endbibitem {}\fi
\ifx \bconflocation  \undefined \def \bconflocation#1{#1}\fi
\ifx \arxivurl  \undefined \def \arxivurl#1{\textsf{#1}}\fi
\csname PreBibitemsHook\endcsname

%%% 1
\bibitem{kantor_trends_2015}
\begin{barticle}
\bauthor{\bsnm{Kantor}, \binits{E.D.}},
\bauthor{\bsnm{Rehm}, \binits{C.D.}},
\bauthor{\bsnm{Haas}, \binits{J.S.}},
\bauthor{\bsnm{Chan}, \binits{A.T.}},
\bauthor{\bsnm{Giovannucci}, \binits{E.L.}}:
\batitle{Trends in {Prescription} {Drug} {Use} {Among} {Adults} in the {United}
  {States} {From} 1999-2012}.
\bjtitle{JAMA}
\bvolume{314}(\bissue{17}),
\bfpage{1818}--\blpage{1830}
(\byear{2015}).
doi:\doiurl{10.1001/jama.2015.13766}.
\bcomment{Publisher: American Medical Association}.
Accessed 2020-08-10
\end{barticle}
\endbibitem

%%% 2
\bibitem{ryu_deep_2018}
\begin{barticle}
\bauthor{\bsnm{Ryu}, \binits{J.Y.}},
\bauthor{\bsnm{Kim}, \binits{H.U.}},
\bauthor{\bsnm{Lee}, \binits{S.Y.}}:
\batitle{Deep learning improves prediction of drug–drug and drug–food
  interactions}.
\bjtitle{Proceedings of the National Academy of Sciences}
\bvolume{115}(\bissue{18}),
\bfpage{4304}--\blpage{4311}
(\byear{2018}).
doi:\doiurl{10.1073/pnas.1803294115}.
\bcomment{Publisher: National Academy of Sciences Section: PNAS Plus}.
Accessed 2020-07-15
\end{barticle}
\endbibitem

%%% 3
\bibitem{ma_genn_2019}
\begin{botherref}
\oauthor{\bsnm{Ma}, \binits{T.}},
\oauthor{\bsnm{Shang}, \binits{J.}},
\oauthor{\bsnm{Xiao}, \binits{C.}},
\oauthor{\bsnm{Sun}, \binits{J.}}:
{GENN}: {Predicting} {Correlated} {Drug}-drug {Interactions} with {Graph}
  {Energy} {Neural} {Networks}.
arXiv:1910.02107 [cs, q-bio, stat]
(2019).
arXiv: 1910.02107.
Accessed 2020-07-15
\end{botherref}
\endbibitem

%%% 4
\bibitem{rohani_drug-drug_2019}
\begin{barticle}
\bauthor{\bsnm{Rohani}, \binits{N.}},
\bauthor{\bsnm{Eslahchi}, \binits{C.}}:
\batitle{Drug-{Drug} {Interaction} {Predicting} by {Neural} {Network} {Using}
  {Integrated} {Similarity}}.
\bjtitle{Scientific Reports}
\bvolume{9}(\bissue{1}),
\bfpage{13645}
(\byear{2019}).
doi:\doiurl{10.1038/s41598-019-50121-3}.
\bcomment{Number: 1 Publisher: Nature Publishing Group}.
Accessed 2020-07-15
\end{barticle}
\endbibitem

%%% 5
\bibitem{rohani_iscmf_2020}
\begin{barticle}
\bauthor{\bsnm{Rohani}, \binits{N.}},
\bauthor{\bsnm{Eslahchi}, \binits{C.}},
\bauthor{\bsnm{Katanforoush}, \binits{A.}}:
\batitle{{ISCMF}: {Integrated} similarity-constrained matrix factorization for
  drug–drug interaction prediction}.
\bjtitle{Network Modeling Analysis in Health Informatics and Bioinformatics}
\bvolume{9}(\bissue{1}),
\bfpage{11}
(\byear{2020}).
doi:\doiurl{10.1007/s13721-019-0215-3}.
Accessed 2020-10-05
\end{barticle}
\endbibitem

%%% 6
\bibitem{vaswani_attention_2017}
\begin{bchapter}
\bauthor{\bsnm{Vaswani}, \binits{A.}},
\bauthor{\bsnm{Shazeer}, \binits{N.}},
\bauthor{\bsnm{Parmar}, \binits{N.}},
\bauthor{\bsnm{Uszkoreit}, \binits{J.}},
\bauthor{\bsnm{Jones}, \binits{L.}},
\bauthor{\bsnm{Gomez}, \binits{A.N.}},
\bauthor{\bsnm{Kaiser}, \binits{L.}},
\bauthor{\bsnm{Polosukhin}, \binits{I.}}:
\bctitle{Attention is {All} you {Need}}.
In: \beditor{\bsnm{Guyon}, \binits{I.}},
\beditor{\bsnm{Luxburg}, \binits{U.V.}},
\beditor{\bsnm{Bengio}, \binits{S.}},
\beditor{\bsnm{Wallach}, \binits{H.}},
\beditor{\bsnm{Fergus}, \binits{R.}},
\beditor{\bsnm{Vishwanathan}, \binits{S.}},
\beditor{\bsnm{Garnett}, \binits{R.}} (eds.)
\bbtitle{Advances in {Neural} {Information} {Processing} {Systems} 30},
pp. \bfpage{5998}--\blpage{6008}.
\bpublisher{Curran Associates, Inc.}, \blocation{???}
(\byear{2017}).
\burl{http://papers.nips.cc/paper/7181-attention-is-all-you-need.pdf}
Accessed 2020-07-15
\end{bchapter}
\endbibitem

%%% 7
\bibitem{zhang_predicting_2017}
\begin{barticle}
\bauthor{\bsnm{Zhang}, \binits{W.}},
\bauthor{\bsnm{Chen}, \binits{Y.}},
\bauthor{\bsnm{Liu}, \binits{F.}},
\bauthor{\bsnm{Luo}, \binits{F.}},
\bauthor{\bsnm{Tian}, \binits{G.}},
\bauthor{\bsnm{Li}, \binits{X.}}:
\batitle{Predicting potential drug-drug interactions by integrating chemical,
  biological, phenotypic and network data}.
\bjtitle{BMC Bioinformatics}
\bvolume{18}(\bissue{1}),
\bfpage{18}
(\byear{2017}).
doi:\doiurl{10.1186/s12859-016-1415-9}.
Accessed 2020-07-15
\end{barticle}
\endbibitem

%%% 8
\bibitem{wan_neodti_2019}
\begin{barticle}
\bauthor{\bsnm{Wan}, \binits{F.}},
\bauthor{\bsnm{Hong}, \binits{L.}},
\bauthor{\bsnm{Xiao}, \binits{A.}},
\bauthor{\bsnm{Jiang}, \binits{T.}},
\bauthor{\bsnm{Zeng}, \binits{J.}}:
\batitle{{NeoDTI}: neural integration of neighbor information from a
  heterogeneous network for discovering new drug–target interactions}.
\bjtitle{Bioinformatics}
\bvolume{35}(\bissue{1}),
\bfpage{104}--\blpage{111}
(\byear{2019}).
doi:\doiurl{10.1093/bioinformatics/bty543}.
\bcomment{Publisher: Oxford Academic}.
Accessed 2020-07-15
\end{barticle}
\endbibitem

%%% 9
\bibitem{gottlieb_indi_2012}
\begin{barticle}
\bauthor{\bsnm{Gottlieb}, \binits{A.}},
\bauthor{\bsnm{Stein}, \binits{G.Y.}},
\bauthor{\bsnm{Oron}, \binits{Y.}},
\bauthor{\bsnm{Ruppin}, \binits{E.}},
\bauthor{\bsnm{Sharan}, \binits{R.}}:
\batitle{{INDI}: a computational framework for inferring drug interactions and
  their associated recommendations}.
\bjtitle{Molecular Systems Biology}
\bvolume{8}(\bissue{1}),
\bfpage{592}
(\byear{2012}).
doi:\doiurl{10.1038/msb.2012.26}.
\bcomment{Publisher: John Wiley \& Sons, Ltd}.
Accessed 2020-07-15
\end{barticle}
\endbibitem

%%% 10
\bibitem{van_laarhoven_gaussian_2011}
\begin{barticle}
\bauthor{\bparticle{van} \bsnm{Laarhoven}, \binits{T.}},
\bauthor{\bsnm{Nabuurs}, \binits{S.B.}},
\bauthor{\bsnm{Marchiori}, \binits{E.}}:
\batitle{Gaussian interaction profile kernels for predicting drug–target
  interaction}.
\bjtitle{Bioinformatics}
\bvolume{27}(\bissue{21}),
\bfpage{3036}--\blpage{3043}
(\byear{2011}).
doi:\doiurl{10.1093/bioinformatics/btr500}.
\bcomment{Publisher: Oxford Academic}.
Accessed 2020-09-23
\end{barticle}
\endbibitem

%%% 11
\bibitem{tatonetti_data-driven_2012}
\begin{barticle}
\bauthor{\bsnm{Tatonetti}, \binits{N.P.}},
\bauthor{\bsnm{Ye}, \binits{P.P.}},
\bauthor{\bsnm{Daneshjou}, \binits{R.}},
\bauthor{\bsnm{Altman}, \binits{R.B.}}:
\batitle{Data-{Driven} {Prediction} of {Drug} {Effects} and {Interactions}}.
\bjtitle{Science Translational Medicine}
\bvolume{4}(\bissue{125}),
\bfpage{125}--\blpage{3112531}
(\byear{2012}).
doi:\doiurl{10.1126/scitranslmed.3003377}.
\bcomment{Publisher: American Association for the Advancement of Science
  Section: Research Article}.
Accessed 2020-07-29
\end{barticle}
\endbibitem

%%% 12
\bibitem{wang_similarity_2014}
\begin{barticle}
\bauthor{\bsnm{Wang}, \binits{B.}},
\bauthor{\bsnm{Mezlini}, \binits{A.M.}},
\bauthor{\bsnm{Demir}, \binits{F.}},
\bauthor{\bsnm{Fiume}, \binits{M.}},
\bauthor{\bsnm{Tu}, \binits{Z.}},
\bauthor{\bsnm{Brudno}, \binits{M.}},
\bauthor{\bsnm{Haibe-Kains}, \binits{B.}},
\bauthor{\bsnm{Goldenberg}, \binits{A.}}:
\batitle{Similarity network fusion for aggregating data types on a genomic
  scale}.
\bjtitle{Nature Methods}
\bvolume{11}(\bissue{3}),
\bfpage{333}--\blpage{337}
(\byear{2014}).
doi:\doiurl{10.1038/nmeth.2810}.
\bcomment{Number: 3 Publisher: Nature Publishing Group}.
Accessed 2020-07-15
\end{barticle}
\endbibitem

%%% 13
\bibitem{paszke_automatic_2017}
\begin{botherref}
\oauthor{\bsnm{Paszke}, \binits{A.}},
\oauthor{\bsnm{Gross}, \binits{S.}},
\oauthor{\bsnm{Chintala}, \binits{S.}},
\oauthor{\bsnm{Chanan}, \binits{G.}},
\oauthor{\bsnm{Yang}, \binits{E.}},
\oauthor{\bsnm{DeVito}, \binits{Z.}},
\oauthor{\bsnm{Lin}, \binits{Z.}},
\oauthor{\bsnm{Desmaison}, \binits{A.}},
\oauthor{\bsnm{Antiga}, \binits{L.}},
\oauthor{\bsnm{Lerer}, \binits{A.}}:
Automatic differentiation in {PyTorch}
(2017).
Accessed 2020-07-29
\end{botherref}
\endbibitem

%%% 14
\bibitem{Chicco2021}
\begin{bbook}
\bauthor{\bsnm{Chicco}, \binits{D.}}:
In: \beditor{\bsnm{Cartwright}, \binits{H.}} (ed.)
\bbtitle{Siamese Neural Networks: An Overview},
pp. \bfpage{73}--\blpage{94}.
\bpublisher{Springer},
\blocation{New York, NY}
(\byear{2021})
\end{bbook}
\endbibitem

%%% 15
\bibitem{He2016}
\begin{bchapter}
\bauthor{\bsnm{He}, \binits{K.}},
\bauthor{\bsnm{Zhang}, \binits{X.}},
\bauthor{\bsnm{Ren}, \binits{S.}},
\bauthor{\bsnm{Sun}, \binits{J.}}:
\bctitle{{Deep residual learning for image recognition}}.
In: \bbtitle{Proceedings of the IEEE Computer Society Conference on Computer
  Vision and Pattern Recognition},
vol. \bseriesno{2016-December},
pp. \bfpage{770}--\blpage{778}.
\bpublisher{IEEE Computer Society}, \blocation{???}
(\byear{2016}).
doi:\doiurl{10.1109/CVPR.2016.90}.
\arxivurl{1512.03385}
\end{bchapter}
\endbibitem

%%% 16
\bibitem{Ba2016}
\begin{botherref}
\oauthor{\bsnm{Ba}, \binits{J.L.}},
\oauthor{\bsnm{Kiros}, \binits{J.R.}},
\oauthor{\bsnm{Hinton}, \binits{G.E.}}:
{Layer Normalization}
(2016).
\arxivurl{1607.06450}
\end{botherref}
\endbibitem

%%% 17
\bibitem{wishart_drugbank_2018}
\begin{barticle}
\bauthor{\bsnm{Wishart}, \binits{D.S.}},
\bauthor{\bsnm{Feunang}, \binits{Y.D.}},
\bauthor{\bsnm{Guo}, \binits{A.C.}},
\bauthor{\bsnm{Lo}, \binits{E.J.}},
\bauthor{\bsnm{Marcu}, \binits{A.}},
\bauthor{\bsnm{Grant}, \binits{J.R.}},
\bauthor{\bsnm{Sajed}, \binits{T.}},
\bauthor{\bsnm{Johnson}, \binits{D.}},
\bauthor{\bsnm{Li}, \binits{C.}},
\bauthor{\bsnm{Sayeeda}, \binits{Z.}},
\bauthor{\bsnm{Assempour}, \binits{N.}},
\bauthor{\bsnm{Iynkkaran}, \binits{I.}},
\bauthor{\bsnm{Liu}, \binits{Y.}},
\bauthor{\bsnm{Maciejewski}, \binits{A.}},
\bauthor{\bsnm{Gale}, \binits{N.}},
\bauthor{\bsnm{Wilson}, \binits{A.}},
\bauthor{\bsnm{Chin}, \binits{L.}},
\bauthor{\bsnm{Cummings}, \binits{R.}},
\bauthor{\bsnm{Le}, \binits{D.}},
\bauthor{\bsnm{Pon}, \binits{A.}},
\bauthor{\bsnm{Knox}, \binits{C.}},
\bauthor{\bsnm{Wilson}, \binits{M.}}:
\batitle{{DrugBank} 5.0: a major update to the {DrugBank} database for 2018}.
\bjtitle{Nucleic Acids Research}
\bvolume{46}(\bissue{D1}),
\bfpage{1074}--\blpage{1082}
(\byear{2018}).
doi:\doiurl{10.1093/nar/gkx1037}.
Accessed 2020-10-14
\end{barticle}
\endbibitem

%%% 18
\bibitem{zhang_label_2015}
\begin{barticle}
\bauthor{\bsnm{Zhang}, \binits{P.}},
\bauthor{\bsnm{Wang}, \binits{F.}},
\bauthor{\bsnm{Hu}, \binits{J.}},
\bauthor{\bsnm{Sorrentino}, \binits{R.}}:
\batitle{Label {Propagation} {Prediction} of {Drug}-{Drug} {Interactions}
  {Based} on {Clinical} {Side} {Effects}}.
\bjtitle{Scientific Reports}
\bvolume{5}(\bissue{1}),
\bfpage{1}--\blpage{10}
(\byear{2015}).
doi:\doiurl{10.1038/srep12339}.
Accessed 2020-10-13
\end{barticle}
\endbibitem

\end{thebibliography}

\newcommand{\BMCxmlcomment}[1]{}

\BMCxmlcomment{

<refgrp>

<bibl id="B1">
  <title><p>Trends in {Prescription} {Drug} {Use} {Among} {Adults} in the
  {United} {States} {From} 1999-2012</p></title>
  <aug>
    <au><snm>Kantor</snm><fnm>ED</fnm></au>
    <au><snm>Rehm</snm><fnm>CD</fnm></au>
    <au><snm>Haas</snm><fnm>JS</fnm></au>
    <au><snm>Chan</snm><fnm>AT</fnm></au>
    <au><snm>Giovannucci</snm><fnm>EL</fnm></au>
  </aug>
  <source>JAMA</source>
  <pubdate>2015</pubdate>
  <volume>314</volume>
  <issue>17</issue>
  <fpage>1818</fpage>
  <lpage>-1830</lpage>
  <url>https://jamanetwork.com/journals/jama/fullarticle/2467552</url>
  <note>Publisher: American Medical Association</note>
</bibl>

<bibl id="B2">
  <title><p>Deep learning improves prediction of drug–drug and drug–food
  interactions</p></title>
  <aug>
    <au><snm>Ryu</snm><fnm>JY</fnm></au>
    <au><snm>Kim</snm><fnm>HU</fnm></au>
    <au><snm>Lee</snm><fnm>SY</fnm></au>
  </aug>
  <source>Proceedings of the National Academy of Sciences</source>
  <pubdate>2018</pubdate>
  <volume>115</volume>
  <issue>18</issue>
  <fpage>E4304</fpage>
  <lpage>-E4311</lpage>
  <url>https://www.pnas.org/content/115/18/E4304</url>
  <note>Publisher: National Academy of Sciences Section: PNAS Plus</note>
</bibl>

<bibl id="B3">
  <title><p>{GENN}: {Predicting} {Correlated} {Drug}-drug {Interactions} with
  {Graph} {Energy} {Neural} {Networks}</p></title>
  <aug>
    <au><snm>Ma</snm><fnm>T</fnm></au>
    <au><snm>Shang</snm><fnm>J</fnm></au>
    <au><snm>Xiao</snm><fnm>C</fnm></au>
    <au><snm>Sun</snm><fnm>J</fnm></au>
  </aug>
  <source>arXiv:1910.02107 [cs, q-bio, stat]</source>
  <pubdate>2019</pubdate>
  <url>http://arxiv.org/abs/1910.02107</url>
  <note>arXiv: 1910.02107</note>
</bibl>

<bibl id="B4">
  <title><p>Drug-{Drug} {Interaction} {Predicting} by {Neural} {Network}
  {Using} {Integrated} {Similarity}</p></title>
  <aug>
    <au><snm>Rohani</snm><fnm>N</fnm></au>
    <au><snm>Eslahchi</snm><fnm>C</fnm></au>
  </aug>
  <source>Scientific Reports</source>
  <pubdate>2019</pubdate>
  <volume>9</volume>
  <issue>1</issue>
  <fpage>13645</fpage>
  <url>https://www.nature.com/articles/s41598-019-50121-3</url>
  <note>Number: 1 Publisher: Nature Publishing Group</note>
</bibl>

<bibl id="B5">
  <title><p>{ISCMF}: {Integrated} similarity-constrained matrix factorization
  for drug–drug interaction prediction</p></title>
  <aug>
    <au><snm>Rohani</snm><fnm>N</fnm></au>
    <au><snm>Eslahchi</snm><fnm>C</fnm></au>
    <au><snm>Katanforoush</snm><fnm>A</fnm></au>
  </aug>
  <source>Network Modeling Analysis in Health Informatics and
  Bioinformatics</source>
  <pubdate>2020</pubdate>
  <volume>9</volume>
  <issue>1</issue>
  <fpage>11</fpage>
  <url>https://doi.org/10.1007/s13721-019-0215-3</url>
</bibl>

<bibl id="B6">
  <title><p>Attention is {All} you {Need}</p></title>
  <aug>
    <au><snm>Vaswani</snm><fnm>A</fnm></au>
    <au><snm>Shazeer</snm><fnm>N</fnm></au>
    <au><snm>Parmar</snm><fnm>N</fnm></au>
    <au><snm>Uszkoreit</snm><fnm>J</fnm></au>
    <au><snm>Jones</snm><fnm>L</fnm></au>
    <au><snm>Gomez</snm><fnm>AN</fnm></au>
    <au><snm>Kaiser</snm><fnm>L</fnm></au>
    <au><snm>Polosukhin</snm><fnm>I</fnm></au>
  </aug>
  <source>Advances in {Neural} {Information} {Processing} {Systems} 30</source>
  <publisher>Curran Associates, Inc.</publisher>
  <editor>Guyon, I. and Luxburg, U. V. and Bengio, S. and Wallach, H. and
  Fergus, R. and Vishwanathan, S. and Garnett, R.</editor>
  <pubdate>2017</pubdate>
  <fpage>5998</fpage>
  <lpage>-6008</lpage>
  <url>http://papers.nips.cc/paper/7181-attention-is-all-you-need.pdf</url>
</bibl>

<bibl id="B7">
  <title><p>Predicting potential drug-drug interactions by integrating
  chemical, biological, phenotypic and network data</p></title>
  <aug>
    <au><snm>Zhang</snm><fnm>W</fnm></au>
    <au><snm>Chen</snm><fnm>Y</fnm></au>
    <au><snm>Liu</snm><fnm>F</fnm></au>
    <au><snm>Luo</snm><fnm>F</fnm></au>
    <au><snm>Tian</snm><fnm>G</fnm></au>
    <au><snm>Li</snm><fnm>X</fnm></au>
  </aug>
  <source>BMC Bioinformatics</source>
  <pubdate>2017</pubdate>
  <volume>18</volume>
  <issue>1</issue>
  <fpage>18</fpage>
  <url>https://doi.org/10.1186/s12859-016-1415-9</url>
</bibl>

<bibl id="B8">
  <title><p>{NeoDTI}: neural integration of neighbor information from a
  heterogeneous network for discovering new drug–target
  interactions</p></title>
  <aug>
    <au><snm>Wan</snm><fnm>F</fnm></au>
    <au><snm>Hong</snm><fnm>L</fnm></au>
    <au><snm>Xiao</snm><fnm>A</fnm></au>
    <au><snm>Jiang</snm><fnm>T</fnm></au>
    <au><snm>Zeng</snm><fnm>J</fnm></au>
  </aug>
  <source>Bioinformatics</source>
  <pubdate>2019</pubdate>
  <volume>35</volume>
  <issue>1</issue>
  <fpage>104</fpage>
  <lpage>-111</lpage>
  <url>https://academic.oup.com/bioinformatics/article/35/1/104/5047760</url>
  <note>Publisher: Oxford Academic</note>
</bibl>

<bibl id="B9">
  <title><p>{INDI}: a computational framework for inferring drug interactions
  and their associated recommendations</p></title>
  <aug>
    <au><snm>Gottlieb</snm><fnm>A</fnm></au>
    <au><snm>Stein</snm><fnm>GY</fnm></au>
    <au><snm>Oron</snm><fnm>Y</fnm></au>
    <au><snm>Ruppin</snm><fnm>E</fnm></au>
    <au><snm>Sharan</snm><fnm>R</fnm></au>
  </aug>
  <source>Molecular Systems Biology</source>
  <pubdate>2012</pubdate>
  <volume>8</volume>
  <issue>1</issue>
  <fpage>592</fpage>
  <url>https://www.embopress.org/doi/full/10.1038/msb.2012.26</url>
  <note>Publisher: John Wiley \& Sons, Ltd</note>
</bibl>

<bibl id="B10">
  <title><p>Gaussian interaction profile kernels for predicting drug–target
  interaction</p></title>
  <aug>
    <au><snm>Laarhoven</snm><fnm>T</fnm></au>
    <au><snm>Nabuurs</snm><fnm>SB</fnm></au>
    <au><snm>Marchiori</snm><fnm>E</fnm></au>
  </aug>
  <source>Bioinformatics</source>
  <pubdate>2011</pubdate>
  <volume>27</volume>
  <issue>21</issue>
  <fpage>3036</fpage>
  <lpage>-3043</lpage>
  <url>https://academic.oup.com/bioinformatics/article/27/21/3036/216840</url>
  <note>Publisher: Oxford Academic</note>
</bibl>

<bibl id="B11">
  <title><p>Data-{Driven} {Prediction} of {Drug} {Effects} and
  {Interactions}</p></title>
  <aug>
    <au><snm>Tatonetti</snm><fnm>NP</fnm></au>
    <au><snm>Ye</snm><fnm>PP</fnm></au>
    <au><snm>Daneshjou</snm><fnm>R</fnm></au>
    <au><snm>Altman</snm><fnm>RB</fnm></au>
  </aug>
  <source>Science Translational Medicine</source>
  <pubdate>2012</pubdate>
  <volume>4</volume>
  <issue>125</issue>
  <fpage>125ra31</fpage>
  <lpage>-125ra31</lpage>
  <url>https://stm.sciencemag.org/content/4/125/125ra31</url>
  <note>Publisher: American Association for the Advancement of Science Section:
  Research Article</note>
</bibl>

<bibl id="B12">
  <title><p>Similarity network fusion for aggregating data types on a genomic
  scale</p></title>
  <aug>
    <au><snm>Wang</snm><fnm>B</fnm></au>
    <au><snm>Mezlini</snm><fnm>AM</fnm></au>
    <au><snm>Demir</snm><fnm>F</fnm></au>
    <au><snm>Fiume</snm><fnm>M</fnm></au>
    <au><snm>Tu</snm><fnm>Z</fnm></au>
    <au><snm>Brudno</snm><fnm>M</fnm></au>
    <au><snm>Haibe Kains</snm><fnm>B</fnm></au>
    <au><snm>Goldenberg</snm><fnm>A</fnm></au>
  </aug>
  <source>Nature Methods</source>
  <pubdate>2014</pubdate>
  <volume>11</volume>
  <issue>3</issue>
  <fpage>333</fpage>
  <lpage>-337</lpage>
  <url>https://www.nature.com/articles/nmeth.2810</url>
  <note>Number: 3 Publisher: Nature Publishing Group</note>
</bibl>

<bibl id="B13">
  <title><p>Automatic differentiation in {PyTorch}</p></title>
  <aug>
    <au><snm>Paszke</snm><fnm>A</fnm></au>
    <au><snm>Gross</snm><fnm>S</fnm></au>
    <au><snm>Chintala</snm><fnm>S</fnm></au>
    <au><snm>Chanan</snm><fnm>G</fnm></au>
    <au><snm>Yang</snm><fnm>E</fnm></au>
    <au><snm>DeVito</snm><fnm>Z</fnm></au>
    <au><snm>Lin</snm><fnm>Z</fnm></au>
    <au><snm>Desmaison</snm><fnm>A</fnm></au>
    <au><snm>Antiga</snm><fnm>L</fnm></au>
    <au><snm>Lerer</snm><fnm>A</fnm></au>
  </aug>
  <pubdate>2017</pubdate>
  <url>https://openreview.net/forum?id=BJJsrmfCZ</url>
</bibl>

<bibl id="B14">
  <title><p>Siamese Neural Networks: An Overview</p></title>
  <aug>
    <au><snm>Chicco</snm><fnm>D</fnm></au>
  </aug>
  <source>Artificial Neural Networks</source>
  <publisher>New York, NY: Springer US</publisher>
  <editor>Cartwright, Hugh</editor>
  <pubdate>2021</pubdate>
  <fpage>73</fpage>
  <lpage>-94</lpage>
</bibl>

<bibl id="B15">
  <title><p>{Deep residual learning for image recognition}</p></title>
  <aug>
    <au><snm>He</snm><fnm>K</fnm></au>
    <au><snm>Zhang</snm><fnm>X</fnm></au>
    <au><snm>Ren</snm><fnm>S</fnm></au>
    <au><snm>Sun</snm><fnm>J</fnm></au>
  </aug>
  <source>Proceedings of the IEEE Computer Society Conference on Computer
  Vision and Pattern Recognition</source>
  <publisher>IEEE Computer Society</publisher>
  <pubdate>2016</pubdate>
  <volume>2016-December</volume>
  <fpage>770</fpage>
  <lpage>-778</lpage>
</bibl>

<bibl id="B16">
  <title><p>{Layer Normalization}</p></title>
  <aug>
    <au><snm>Ba</snm><fnm>JL</fnm></au>
    <au><snm>Kiros</snm><fnm>JR</fnm></au>
    <au><snm>Hinton</snm><fnm>GE</fnm></au>
  </aug>
  <pubdate>2016</pubdate>
  <url>http://arxiv.org/abs/1607.06450</url>
</bibl>

<bibl id="B17">
  <title><p>{DrugBank} 5.0: a major update to the {DrugBank} database for
  2018</p></title>
  <aug>
    <au><snm>Wishart</snm><fnm>DS</fnm></au>
    <au><snm>Feunang</snm><fnm>YD</fnm></au>
    <au><snm>Guo</snm><fnm>AC</fnm></au>
    <au><snm>Lo</snm><fnm>EJ</fnm></au>
    <au><snm>Marcu</snm><fnm>A</fnm></au>
    <au><snm>Grant</snm><fnm>JR</fnm></au>
    <au><snm>Sajed</snm><fnm>T</fnm></au>
    <au><snm>Johnson</snm><fnm>D</fnm></au>
    <au><snm>Li</snm><fnm>C</fnm></au>
    <au><snm>Sayeeda</snm><fnm>Z</fnm></au>
    <au><snm>Assempour</snm><fnm>N</fnm></au>
    <au><snm>Iynkkaran</snm><fnm>I</fnm></au>
    <au><snm>Liu</snm><fnm>Y</fnm></au>
    <au><snm>Maciejewski</snm><fnm>A</fnm></au>
    <au><snm>Gale</snm><fnm>N</fnm></au>
    <au><snm>Wilson</snm><fnm>A</fnm></au>
    <au><snm>Chin</snm><fnm>L</fnm></au>
    <au><snm>Cummings</snm><fnm>R</fnm></au>
    <au><snm>Le</snm><fnm>D</fnm></au>
    <au><snm>Pon</snm><fnm>A</fnm></au>
    <au><snm>Knox</snm><fnm>C</fnm></au>
    <au><snm>Wilson</snm><fnm>M</fnm></au>
  </aug>
  <source>Nucleic Acids Research</source>
  <pubdate>2018</pubdate>
  <volume>46</volume>
  <issue>D1</issue>
  <fpage>D1074</fpage>
  <lpage>-D1082</lpage>
  <url>http://academic.oup.com/nar/article/46/D1/D1074/4602867</url>
</bibl>

<bibl id="B18">
  <title><p>Label {Propagation} {Prediction} of {Drug}-{Drug} {Interactions}
  {Based} on {Clinical} {Side} {Effects}</p></title>
  <aug>
    <au><snm>Zhang</snm><fnm>P</fnm></au>
    <au><snm>Wang</snm><fnm>F</fnm></au>
    <au><snm>Hu</snm><fnm>J</fnm></au>
    <au><snm>Sorrentino</snm><fnm>R</fnm></au>
  </aug>
  <source>Scientific Reports</source>
  <pubdate>2015</pubdate>
  <volume>5</volume>
  <issue>1</issue>
  <fpage>1</fpage>
  <lpage>-10</lpage>
  <url>https://www.nature.com/articles/srep12339</url>
</bibl>

</refgrp>
} % end of \BMCxmlcomment

\end{document}